# Modeling Fractional Polytropic Gas Spheres Using Artificial Neural Network

Mohamed I. Nouh[1], Yosry A. Azzam[1] and Emad A.-B. Abdel-Salam[2]

[1]Astronomy Department, National Research Institute of Astronomy and Geophysics (NRIAG), 11421 Helwan, Cairo, Egypt

[2]Department of Mathematics, Faculty of Science, New Valley University, El-Kharja 72511, Egypt

e-mail: mohamed.nouh@nriag.sci.eg

**Abstract:**

Lane-Emden differential equations describe different physical and astrophysical phenomena that include forms of stellar structure, isothermal gas spheres, gas spherical cloud thermal history, and thermionic currents. This paper presents a computational approach to solve the problems related to fractional Lane-Emden differential equations based on neural networks. Such a solution will help solve the fractional polytropic gas spheres problems which have different applications in physics, astrophysics, engineering, and several real-life issues. We used Artificial Neural Network (ANN) framework in its feedforward back propagation learning scheme. The efficiency and accuracy of the presented algorithm are checked by testing it on four fractional Lane-Emden equations and compared with the exact solutions for the polytopic indices n=0,1,5 and those of the series expansions for the polytropic index n=3. The results we obtained prove that using the ANN method is feasible, accurate, and may outperform other methods.

1. Introduction

The nonlinear Lane-Emden differential equations (polytropic and isothermal) have a singularity at origin and possess only exact solutions for the polytropic index n=0, 1, and 5 [1-2]. In astrophysics, these equations could be used to model many problems such that; spherical cloud of gas, stellar structure, and galactic structure. There are several methods proposed to solve the integer version of these equations i.e. homotopy perturbation method [3], variational iteration method [4], Sinc-Collocation method [5], an implicit series solution [6], accelerated series solution [7-8] and Adomian decomposition method [9].

Recently and due to their wide applications in science and engineering, analytical solutions to the fractional nonlinear differential equations acquired great attention, [10-11]. The Newtonian stellar polytrope's fractional version was investigated by [12] for the fractional white dwarf model, [13] for the incompressible gas sphere, [14] for the fractional isothermal gas sphere. Nouh and Abdel-Salam [15] and [16] constructed fractional polytropic stellar models for white dwarfs with n=1.5 and solar type stars n=3. For n =3, [15-16] showed that solar-like stars could be modeled by dividing the interior into two layers with different fractional parameters.

Neural networks (NNs) have acquired a solid role in many areas of human activity over the past decades and have found application in a wide range of scientific issues, including astronomy, geology, geophysics, and the environmental sciences [17-21]. It was commonly used in the areas of pattern recognition, data classification, prediction, function approximation, signal processing, medical diagnosis, modeling, and control, etc. [22-26]. The Artificial Neural Networks (ANN) try to imitate the biological brain mathematically, in which linear or nonlinear processes information and neuron models are connected in a parallel and distributed style. ANN performs computations at a much higher speed because of its massively parallel nature. They have the capabilities of learning and self-organization that can memorize and pick up a mapping between an input and an output vector space and synthesize an associative memory which recovers the correct output when the input is introduced and generalizes when new inputs are introduced [27].

Because of their excellent properties of fault tolerance, self-learning, adaptivity, nonlinearity, ANNs are used mostly nowadays for function approximation in numerical models [28]. The finite-time and adaptive finite-time synchronization principle in graph theory perspectives have been investigated under two different control strategies by Pratab et al.[29]. Zhou et al. [30] consider the finite-time synchronization of dynamic networks with nonlinear coupling strength and stochastic perturbations by using intermittent control. Zhang et al. [31] investigate a cluster of limitless, delayed neural networks spread.

Besides, ANNs have been extensively used to solve linear and nonlinear differential equation related problems with integer or non-integer derivative and used various paradigms for ANN architecture [32-39]. Besides, Ahmad et al. [40] computed the solution of Lane–Emden type equations by the use of artificial neural networks (ANNs). Based on active-set (AS), interior-point (IP), and sequential quadratic programming (SQP) algorithms, local optimization procedures have been used in this research to optimize the energy functions. Jalab et al. [41] introduced a neural network based numerical method, for solving the integer and fractional Lane-Emden type equations.

In the present paper, we will formulate the fractional Lane-Emden equation of the polytropic gas sphere and solve it analytically and train the ANN algorithm using tables of the fractional Emden functions and mass-radius relation computed by means of the accelerated series expansion method. For the numerical simulation, we use the ordinary feed-forward neural network to approximate the solution of fractional Lane-Emden equation of the polytropic gas sphere and mass-radius relation which is proved to have more benefits when compared to other computational methods. The structure we used is a three-layer feed-forward neural network that is trained using the back-propagation learning algorithm based on the gradient descent rule. The rest of the paper is organized as follows: Section 2 deals with the principle of the conformable fractional derivatives. The derivation of the polytropic Lane-Emden equation is performed in section 3. Section 4 is devoted to the neural network algorithm. In section 5, the results are outlined. We give the conclusion reached in section 6.

## 2. Conformable fractional derivatives

Different definitions of fractional derivatives exist. Examples include Riemann–Liouville, Kolwankar–Gangal, Caputo, modified Riemann–Liouville, Cresson's, and Chen's fractal derivatives, [42] and [10]. The conformable fractional derivative (CFD) introduced by Khalil et al. [43] used the limits in the form:

$$D^\alpha f(t) = \lim_{\varepsilon \to 0} \frac{f(t + \varepsilon t^{1-\alpha}) - f(t)}{\varepsilon} \quad \forall t > 0, \, \alpha \in (0,1], \tag{1}$$

$$f^{(\alpha)}(0) = \lim_{t \to 0^+} f^{(\alpha)}(t). \tag{2}$$

Here $f^{(\alpha)}(0)$ is not defined. This fractional derivative reduces to the ordinary derivative when $\alpha = 1$. The following properties are found in the conformable fractional derivative:

$$D^\alpha t^p = p t^{p-\alpha}, \quad p \in \mathbb{R}, \quad D^\alpha c = 0, \quad \forall f(t) = c, \tag{3}$$

$$D^\alpha (a f + b g) = a D^\alpha f + b D^\alpha g, \quad \forall a,b \in \mathbb{R}, \tag{4}$$

$$D^\alpha (f g) = f D^\alpha g + f D^\alpha g, \tag{5}$$

$$D^\alpha f(g) = \frac{df}{dg} D^\alpha g, \qquad D^\alpha f(t) = t^{1-\alpha} \frac{df}{dg}, \tag{6}$$

where $f, g$ are two $\alpha-$differentiable functions and $c$ constant is an arbitrary constant. Equations (5) to (6) are demonstrated by [43]. The corresponding fractional derivative of certain functions could be given by:

$$\begin{aligned} D^\alpha e^{ct} &= c t^{1-\alpha} e^{ct}, \quad D^\alpha \sin(ct) = c t^{1-\alpha} \cos(ct), \quad D^\alpha \cos(ct) = -c t^{1-\alpha} \sin(ct), \\ D^\alpha e^{ct^\alpha} &= c \alpha e^{ct^\alpha}, \quad D^\alpha \sin(ct^\alpha) = c \alpha \cos(ct^\alpha), \quad D^\alpha \cos(ct^\alpha) = -c \alpha \sin(ct^\alpha). \end{aligned} \tag{7}$$

### 3. Fractional Polytropic Gas Spheres

The polytropic equation of state has the form

$$p = K \rho^\gamma, \quad \gamma = 1 + \frac{1}{n}. \tag{8}$$

Where $K$ is the pressure constant and $n$ is the polytropic index A self-gravitating object's equilibrium structure is derived from hydrostatic equations. The simplest case is a spherical, non-rotating, static configuration in which all macroscopic properties for a given equation of state are parameterized by a single parameter, e.g. central density. The equation representing the conformable fractional form for mass conservation and hydrostatic equilibrium is given by:

$$D_r^\alpha M = 4\pi r^{2\alpha} \rho, \tag{9}$$

and

$$D_r^\alpha P = -\frac{GM}{r^{2\alpha}} \rho. \tag{10}$$

Rearrange Equation (10) we get

$$\frac{r^{2\alpha}}{\rho} D_r^\alpha P = -GM, \tag{11}$$

By conducting the first fractional derivative of Equation (11) we will get

$$D_r^\alpha \left( \frac{r^{2\alpha}}{\rho} D_r^\alpha P \right) = -G D_r^\alpha M ,\qquad(12)$$

Combining Equations (11) and (12) we get

$$D_r^\alpha \left( \frac{r^{2\alpha}}{\rho} D_r^\alpha P \right) = -4\pi G r^{2\alpha} \rho ,\qquad(13)$$

or

$$\frac{1}{r^{2\alpha}} D_r^\alpha \left( \frac{r^{2\alpha}}{\rho} D_r^\alpha P \right) = -4\pi G \rho \qquad(14)$$

Now, by defining the u function (Emden function), which is a dimensionless function, as:

$$\rho = \rho_c u^n ,\qquad(15)$$

where $\rho$ and $\rho_c$ are the density and central density respectively. The dimensionless variable $x$ could be written as

$$x^\alpha = \frac{r^\alpha}{a} .\qquad(16)$$

Inserting Equations (8) and (12) in Equation (14) we get

$$\frac{1}{(ax^\alpha)^2} \frac{d^\alpha}{d(ax^\alpha)} \left( \frac{(ax^\alpha)^2}{\rho_c u^n} \frac{d^\alpha (K\rho^\gamma)}{d(ax^\alpha)} \right) = -4\pi G \rho_c u^n ,\qquad(17)$$

$$\frac{K}{(ax^\alpha)^2} \frac{d^\alpha}{d(ax^\alpha)} \left( \frac{(ax^\alpha)^2}{\rho_c u^n} \frac{d^\alpha (\rho_c u^n)^{1+\frac{1}{n}}}{d(ax^\alpha)} \right) = -4\pi G \rho_c u^n .\qquad(18)$$

The Emden's fractional derivative $u$ could be written as

$$\frac{d^\alpha}{dx^\alpha} u^{n+1} = (n+1) u^n \frac{d^\alpha u}{dx^\alpha} .\qquad(19)$$

Inserting Equation (19) in Equation (18) we get

$$\frac{K}{a^2 x^{2\alpha}} \frac{d^\alpha}{d x^\alpha} \left( \frac{(n+1)x^{2\alpha} \rho_c^{1+\frac{1}{n}} u^n}{\rho_c u^n} \frac{d^\alpha u}{d x^\alpha} \right) = -4\pi G \rho_c u^n , \qquad (20)$$

or

$$\frac{K}{a^2 x^{2\alpha}} D_X^\alpha \left( \frac{(n+1)x^{2\alpha} \rho_c^{1+\frac{1}{n}} u^n}{\rho_c u^n} D_X^\alpha u \right) = -4\pi G \rho_c u^n , \qquad (21)$$

rearrange

$$\frac{K(n+1)\rho_c^{\frac{1}{n}-1}}{4\pi G a^2} \frac{1}{x^{2\alpha}} D_X^\alpha \left( x^{2\alpha} D_X^\alpha u \right) = -u^n . \qquad (22)$$

Now by taking

$$a^2 = \frac{K(n+1)\rho_c^{\frac{1}{n}-1}}{4\pi G}, \qquad (23)$$

then the fractional form of Lane-Emden equation is given by:

$$\frac{1}{x^{2\alpha}} D_X^\alpha \left( x^{2\alpha} D_X^\alpha u \right) = -u^n . \qquad (24)$$

with the initial conditions:

$$u(0) = 1, \qquad D_x^\alpha u(0) = 0 \qquad (25)$$

where $u = u(x)$, is the Emden function and $0 < \alpha \leq 1$.

Assume the transform $X = x^\alpha$, the Emden function will take the form [16]

$$u(X) = \sum_{m=0}^{\infty} A_m X^m , \qquad (26)$$

the series expansion coefficients are written as

$$A_{k+2} = -\frac{Q_k}{\alpha^2(k+2)(k+3)}, \quad \forall\, k \geq 2 \tag{27}$$

and

$$Q_m = \frac{1}{m!A_0} \sum_{i=1}^{m} (m-1)!(in - m + i) A_i Q_{m-i}, \quad \forall\, m \geq 1, \tag{28}$$

We get the series coefficients of the integer LEE by putting $\alpha = 1$ in Equations (27) and (28). If we insert $k = 0, 1, 2, 3$ in Equations (27) and (28) we will get:

$$A_0 = 1,\ A_1 = 0,\quad Q_0 = 1,\ Q_1 = 0,\ A_2 = -\frac{1}{6\alpha^2},\ A_3 = 0,\ A_4 = \frac{n}{120\alpha^4},\ A_5 = 0.$$

Consequently, the series solution at $\alpha = 1$ is reduced to the integer version of LEE [7] as

$$u_n(x) = 1 - \frac{1}{6}x + \frac{n}{120}x^2 - \ldots\ldots\ldots$$

For n = 0, 1 and 5, the exact solutions are given by

$$u(x) = 1 - \frac{1}{6}\left(\frac{x^\alpha}{\alpha}\right)^2, \tag{29}$$

$$u(x) = \left(\frac{x^\alpha}{\alpha}\right)^{-1} \sin\left(\frac{x^\alpha}{\alpha}\right), \tag{30}$$

and

$$u(x) = \left(1 + \frac{1}{3}\left(\frac{x^\alpha}{\alpha}\right)^2\right)^{-\frac{1}{2}}. \tag{31}$$

The radius, mass, and density of the polytrope could be given by [15] and [16]

$$M(x^\alpha) = 4\pi \left[\frac{K(n+1)}{4\pi G}\right]^{\frac{3}{2}} \rho_c^{\frac{3-n}{2n}} \left[-\left(x^{2\alpha}\frac{d^\alpha u}{d\,x^\alpha}\right)\right]_{x=x_1}, \tag{32}$$

$$R^\alpha = \left[\frac{K(n+1)}{4\pi G}\right]^{\frac{1}{2}} \rho_c^{\frac{1-n}{2n}} x_1^\alpha. \tag{33}$$

## 4. Neural network algorithm

### 4.1 Mathematical modeling of the problem

The neural network architecture proposed to simulate the fractional Lane-Emden equation is as shown in Figure 1. The fractional Lane-Emden equation could be written as

$$D_X^{\alpha\alpha}u + \frac{1}{x^{2\alpha}}D_X^{\alpha}u = -u^n. \tag{34}$$

To obtain a neural network solution along with the initial conditions $u(0)=1$ and $D_x^{\alpha}u(0)=0$, we perform the following [44]:

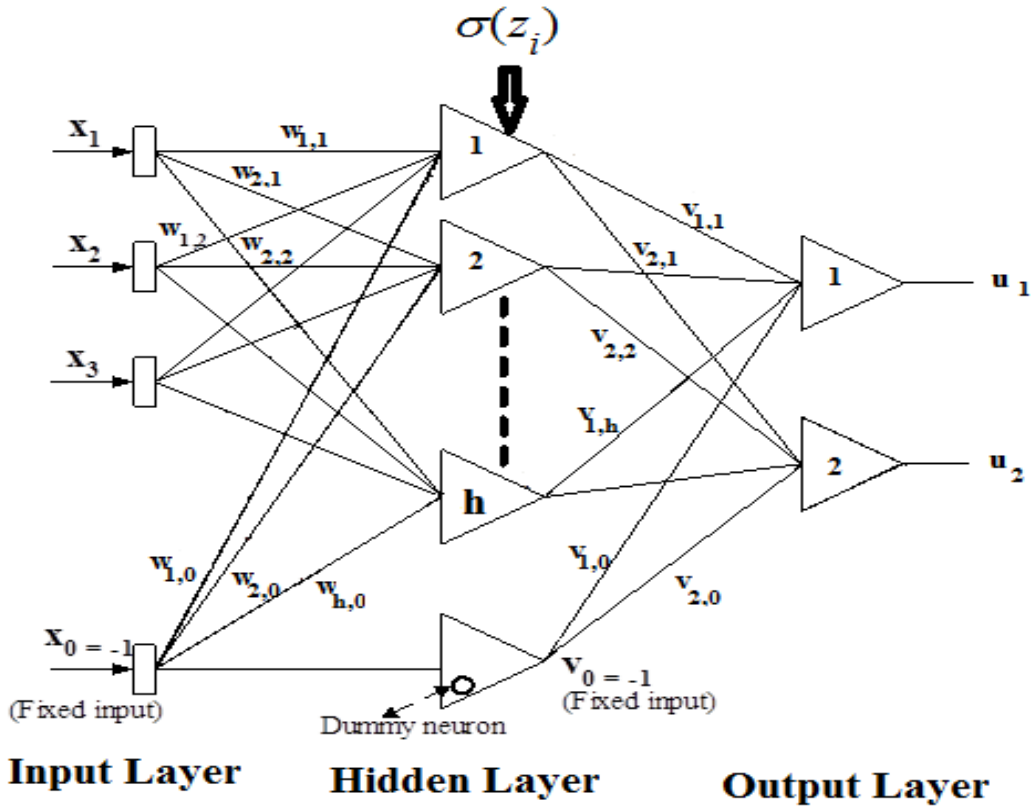

**Figure 1. ANN architecture proposed to simulate the fractional Lane-Emden equation.**

First, we consider $u_t(x,p)$ as the approximate solution of the neural network for Equation (34) and can be formulated to be of the following form:

$$u_t(x,p) = A(x) + f(x, N(x,p)), \tag{35}$$

where the first term satisfies the initial or boundary values and the second term represents feed forward neural network with input vector $x$ and $p$ is the corresponding vector of adjustable weight parameters. The neural network output $N(x,p)$ is given by

$$N(x,p) = \sum_{i=1}^{H} v_i \sigma(z_i), \qquad (36)$$

where $z_j = \sum_{i=1}^{n} w_{ij} x_j + \beta_i$ and $w_{ij}$ denotes the weight from the input unit $j$ to the hidden unit $i$, $v_i$ represents weight from the hidden unit $i$ to the output, $\beta_i$ is the bias of the $i$th hidden unit, and $\sigma(z_i)$ is the sigmoid activation function that has the form $\sigma(x) = \dfrac{1}{1+e^{-x}}$.

Now the derivative of networks output $N$ for input vector $x_j$ is

$$D_{x_j}^{\alpha} N(x,p) = D_{x_j}^{\alpha}\left(\sum_{i=1}^{H} v_i \sigma\left(z_i = \sum_{i=1}^{n} w_{ij} x_j + \beta_i\right)\right) = \sum_{i=1}^{h} v_i w_{ij} \sigma^{(\alpha)}, \qquad \sigma^{(\alpha)} = D_x^{\alpha} \sigma(x), \qquad (37)$$

Similarly, the $n^{th}$ fractional derivative of $N$, Equation (36), is

$$D_{x_j}^{\alpha \ \overset{n\ times}{\ldots} \alpha} N(x,p) = \sum_{i=1}^{n} v_i \ P_i \ \sigma_i^{(n\alpha)}, \qquad , P_i = \prod_{k=1}^{n} w_{ik}^{\alpha_k}, \qquad \sigma_i = \sigma(z_i), \qquad (38)$$

Hence, the proposed approximate solution for the fractional Lane-Emden equation is given as

$$u_t(x,p) = 1 + x\, N(x,p), \qquad (39)$$

which satisfies the initial conditions as:

$$u_t(0,p) = 1 + 0.N(0,p) = 1, \qquad (40)$$

and

$$D_x^{\alpha} u_t(x,p) = x^{1-\alpha} N(x,p) + x\, D_x^{\alpha} N(x,p), \qquad (41)$$

so

$$D_x^{\alpha} u_t(0,p) = (0)^{1-\alpha} N(x,p) + 0.D_x^{\alpha} N(x,p) = 0, \qquad (42)$$

### 4.2 ANN gradient computations and parameter updating

Now, if we considered the approximate solution represented by Equation (39), the problem is converted into an unconstrained optimization problem and the error quantity to be minimized can be given by

$$E(x) = \sum_i \left\{ D_x^{\alpha\alpha} u_t(x_i, p) + \frac{2}{x} D_x^{\alpha} u_t(x_i, p) - f(x_i, u_t(x_i, p)) \right\}^2, \tag{43}$$

where

$$D_x^{\alpha} u_t(x, p) = x^{1-\alpha} N(x, p) + x D_x^{\alpha} N(x, p), \tag{44}$$

and

$$D_x^{\alpha\alpha} u_t(x, p) = (1-\alpha)x^{1-2\alpha} N(x, p) + 2x^{1-\alpha} D_x^{\alpha} N(x, p) + x D_x^{\alpha\alpha} N(x, p), \tag{45}$$

where $D_x^{\alpha} N(x, p)$ and $D_x^{\alpha\alpha} N(x, p)$ is given by Equations (37-38).

For network parameter updating, we compute the fractional derivative of the neural network for input as well as for parameters of the network and train the neural network for the optimized value of parameters. Once the network is trained set up the network with optimized network parameters and compute $u_t(x, p)$ from $u_t(x, p) = 1 + x N(x, p)$.

The conformable fractional derivative with respect to any of its inputs is equivalent to a feed-forward neural network $N$ with one hidden layer, having the same values for the weights $w_{ij}$ and thresholds $\beta_i$ and with each weight $v_i$ being replaced with $v_i P_i$ where $P_i = \prod_{k=1}^{n} w_{ik}^{\alpha_k}$. Moreover, the transfer function of each hidden unit is replaced with the n$^{th}$ order fractional derivative of the sigmoid function. Therefore, the conformable fractional gradient of $N$ with respect to the parameters of the original network can be obtained as:

$$\begin{aligned} D_{v_i}^{\alpha} N &= P_i \, \sigma_i^{(n\alpha)} \\ D_{\beta_i}^{\alpha} N &= v_i P_i \, \sigma_i^{((n+1)\alpha)} \\ D_{w_{ij}}^{\alpha} N &= x_i v_i P_i \, \sigma_i^{((n+1)\alpha)} + v_i \alpha_j w_{ij}^{1-\alpha_j} \left( \prod_{k=1, k \neq j} w_{ik}^{\alpha_k} \right) \sigma_i^{(n\alpha)} \end{aligned} \tag{46}$$

The network parameters updating rule can be given as,

$$v_i(x+1) = v_i(x) + a D_{v_i}^{\alpha} N, \tag{47}$$

$$\beta_i(x+1) = \beta_i(x) + b D_{\beta_i}^{\alpha} N, \tag{48}$$

$$w_{ij}(x+1) = w_{ij}(x) + c D_{w_{ij}}^{\alpha} N, \tag{49}$$

where $a$, $b$, $c$ are learning rates, $i = 1, 2, \ldots, n$, and $j = 1, 2, \ldots, h$.

In ANN, the main processing unit which can carry out localized information and can process a local memory is the neuron. The net input ($z$) at each neuron is calculated by adding the weights it receives to get a weighted sum of those inputs and add it with a bias ($\beta$). Then the net input ($z$) is passed through an activation function, resulting in the output of the neuron $u_j$ (as is shown in Figure 1).

### 4.3. Back-propagation learning algorithm

Different algorithms for training the neural network are found in the literature. The traditional and the most famous algorithm is the steepest descent algorithm which is also known as the error backpropagation (BP) algorithm [45]. The BP training algorithm is a gradient algorithm designed to minimize the mean square error between the actual output of a feed-forward net and the desired output. It requires continuously differentiable non-linearity. Although the convergence rate of this algorithm is slow [46], its stability is high compared to other training algorithms [47]. The mathematics of the gradient algorithm has to guarantee that a particular node has to be adjusted in direct proportion to the error in the units to which it is connected. The BP algorithm performs the steepest descent on a surface in a weight space whose height at any point in weight space is equal to the error measure. The error function which has to reduce can be written in the following form:

$$E = \frac{1}{|D|} \sum_{x \in D} F(x, U)^2 \tag{50}$$

Here, $F$ is some signed error measure, $D$ is a set of training patterns at which error is to be evaluated and $U$ represents the neural network output [48]. We can define the state of the unit to be the weighted sum of the output of the previous layer as [49]:

$$S_{pj} = \sum_i W_{ji} O_{pi} \tag{51}$$

The output,

$$O_{pj} = f_j(S_{pj}) \tag{52}$$

uses the sigmoid function. To get the correct generalization of the delta rule, $W_{ji}$ is set as:

$$\Delta_p W_{ji} \propto -\frac{\partial E_p}{\partial W_{ji}} \tag{53}$$

It may be useful to see this derivative to be resulting from the product of two parts: one part reflecting the change in the net input to the unit and the other part representing the effect of changing a particular weight on the net input. Thus, we can write

$$\frac{\partial E_p}{\partial W_{ji}} = \frac{\partial E_p}{\partial S_{pj}} \frac{\partial S_{pj}}{\partial W_{ji}} \tag{54}$$

From (51), we can see that the second factor is:

$$\frac{\partial S_{pj}}{\partial W_{ji}} = \frac{\partial}{\partial W_{ji}} \sum_k W_{jk} O_{pk} = O_{pi} \tag{55}$$

Now, we can define:

$$\delta_{pj} = -\frac{\partial E_p}{\partial S_{pj}} \tag{55}$$

Equation (54) thus has the equivalent form

$$-\frac{\partial E_p}{\partial W_{ji}} = \delta_{pj} O_{pi} \tag{56}$$

This tells that to implement a gradient descent in *E*, we should make the weight changes according to:

$$\Delta_p W_{ji} = \eta \delta_{pj} O_{pi} \tag{57}$$

Where η is the learning rate factor.

It is interesting to see that there is a simple recursive computation of these δ's that can be implemented by propagating an error signal back through the network. To compute equation (55), the chain rule is applied to write this partial derivative as the product of two factors, one factor reflecting the change in error as a function of the output of the unit, and the other one reflecting the change in the output as a function of changes in the input,

$$\delta_{pj} = -\frac{\partial E_p}{\partial S_{pj}} = -\frac{\partial E_p}{\partial O_{pj}} \frac{\partial O_{pj}}{\partial S_{pj}} \tag{58}$$

By (52), we can see that

$$\frac{\partial O_{pj}}{\partial S_{pj}} = f_j^{'}(S_{pj}) \tag{59}$$

Which is the derivative of the compressing function $f_j$ for the *jth* unit, evaluated at the net input $S_{pj}$ to that unit. To compute the first factor, there are two cases. First, assume that the unit $U_i$ is an output unit of the network. In this case, it follows from the definition of $E_p$ that:

$$\frac{\partial E_p}{\partial O_{pj}} = -(T_{pj} - O_{pj}) \tag{60}$$

Substituting for the two factors in (58), we can get:

$$\delta_{pj} = (T_{pj} - O_{pj}) f_j'(S_{pj}) \tag{61}$$

for any output unit $U_j$. If $U_j$ is not an output unit, the chain rule is used to write

$$\sum_k \frac{\partial E_p}{\partial S_{pk}} \frac{\partial S_{pk}}{\partial O_{pj}} = \sum_k \frac{\partial E_p}{\partial S_{pk}} \frac{\partial}{\partial O_{pj}} \sum_i W_{ki} O_{pi} = \sum_k \frac{\partial E_p}{\partial S_{pk}} W_{kj} = -\sum_k \delta_{pk} W_{kj} \tag{62}$$

In this case, substituting for the two factors in (58) yields

$$\delta_{pj} = f_j'(S_{pj}) \sum_k \delta_{pk} W_{pj} \tag{63}$$

Whenever $U_j$ is not an output unit. Equations (61) and (63) give a recursive procedure for computing the $\delta$'s for all units in the network, which are then used to compute the weight changes in the network according to (57). Figure 2 shows a flow chart of a back-propagation off-line learning algorithm [49]. As is seen, a comparison of the output $u_j$ at the output layer with the target output $t_j$ is implemented using an error function that has the following form:

$$\delta_j = u_j(t_j - u_j)(1 - u_j). \tag{64}$$

For the hidden layer, the error function takes the form:

$$\delta_j = u_j(1 - u_j) \sum_k \delta_k w_k. \tag{65}$$

where $\delta_j$ is the output layer error term, and $w_k$ is the weight between the hidden and output layers. To update the weight of each connection, the error is replicated backward from the output layer to the input layer as follows:

$$w_{ji}(t+1) = w_{ji}(t) + \eta \delta_j u_j + \gamma(w_{ji}(t) - w_{ji}(t-1)) \tag{66}$$

Learning rate $\eta$ has to be chosen such that it is neither very small leading to a slow rate of convergence nor too large leading to overshooting. The last term in Equation (66) is called the momentum term and is added with the momentum constant $\gamma$ to speed up the convergence of the back-propagation learning algorithm error and to help in kicking the changes over local increases in the energy function and pushing the weights to follow the overall downhill direction [50]. This term has the effect of adding a fraction of the most recent weight adjustment to the current weight adjustments. Both η and $\gamma$ terms are assigned at the beginning of the training phase and decide the network stability and speed [21], [27].

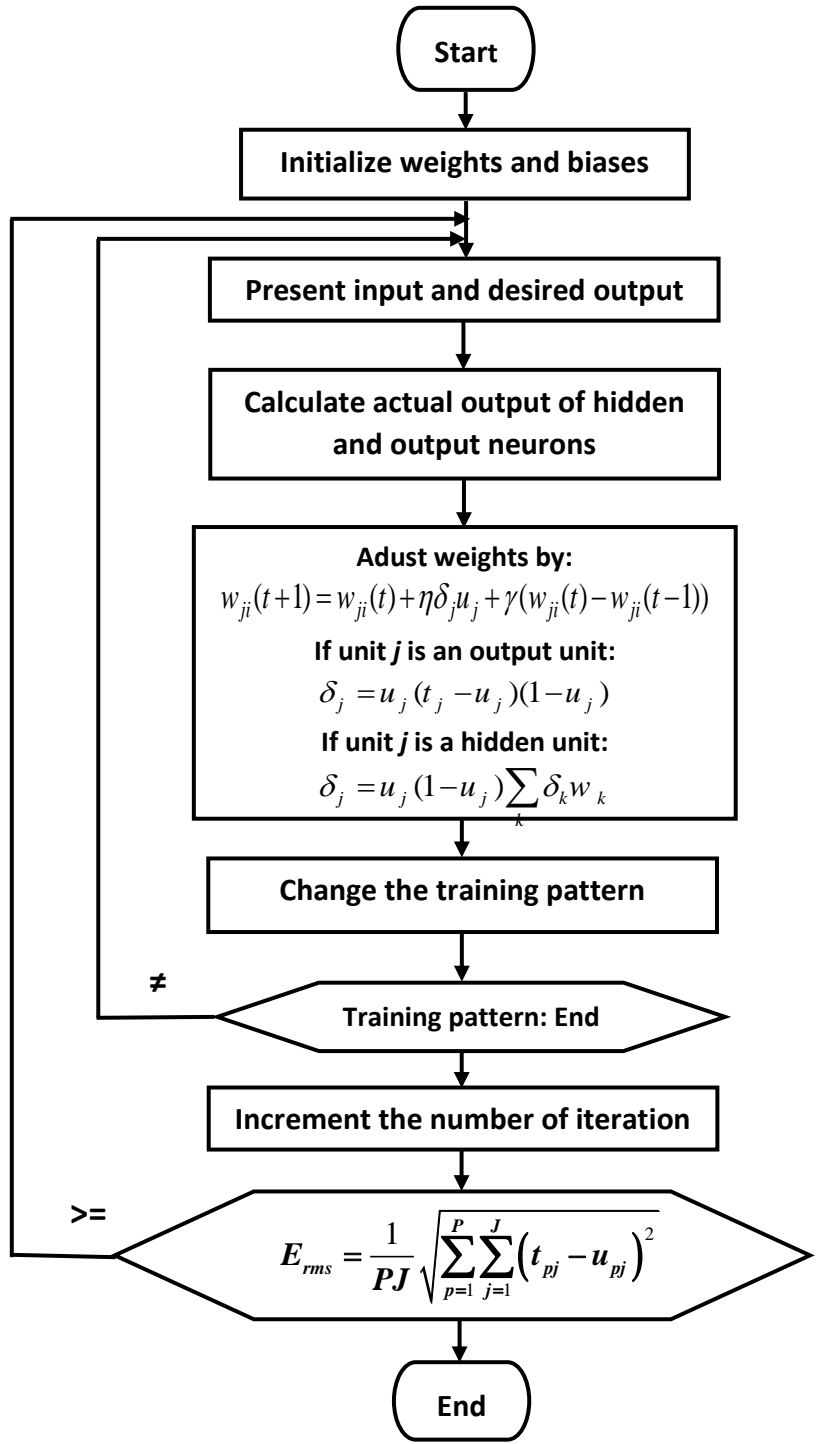

**Figure 2. Flowchart of ANN back-propagation off-line training algorithm**

For each input pattern, the process is repeated until the network output error is reduced to a pre-assigned threshold value. The final weights are frozen and used to obtain the exact fractional values of Lane-Emden differential equations during the test session. To assess the success and quality of the training, an error is calculated for the whole batch of training patterns using the root-mean-square normalized error which is defined as:

$$E_{rms} = \frac{1}{PJ} \sqrt{\sum_{p=1}^{P}\sum_{j=1}^{J}\left(t_{pj} - u_{pj}\right)^2} \qquad (67)$$

where $P$ is the number of training patterns, $J$ is the number of output units, $t_{pj}$ is target output at unit $j$, and $u_{pj}$ is the actual output at the same unit $j$. An error of zero would indicate that all the output patterns calculated by the ANN match the expected values perfectly and that the ANN is well trained. Internal unit thresholds are adapted similarly by assuming they are connection weights on links from auxiliary constant-valued input.

We have programmed the previous algorithms using C++ programming language running on Windows 7 of a CORE i7 PC.

## 5. Numerical results and discussion

### 5.1 Preparation of the input data

To prepare the input data for the training procedure of the proposed ANN algorithm, we compute Emden functions and the physical parameters at various polytropic indices and fractional parameters. What concerns us in the present calculations is the polytropes having exact solutions, namely, polytropes with n=0, 1, 5. Besides, we study the polytropic mass-radius relation for normal stars with n=3. As a result, we will have four versions of the fractional Lane-Emden differential equation being extracted from Equation (24) which represents the different polytropic indices under study. These equations are:

$$D_X^{\alpha\alpha} u + \frac{1}{x^{2\alpha}} D_X^{\alpha} u = -1, \qquad (68)$$

$$D_X^{\alpha\alpha} u + \frac{1}{x^{2\alpha}} D_X^{\alpha} u = -u, \qquad (69)$$

$$D_X^{\alpha\alpha}u + \frac{1}{x^{2\alpha}}D_X^{\alpha}u = -u^3, \tag{70}$$

$$D_X^{\alpha\alpha}u + \frac{1}{x^{2\alpha}}D_X^{\alpha}u = -u^5, \tag{71}$$

for the polytropic indices n=0, 1, 3, 5 respectively.

The general series solutions of the above four equations (Equation (26)), and by implementing the two recurrence relations, Equations (27-28), could be written as

$$u_n(x) = 1 - \frac{1}{6\alpha}x + \frac{n}{120\alpha^2}x^2 - \cdots\cdots \tag{72}$$

The series presented by Equation (72) not converge to the surface of the polytope, so it may be used only to model the region near the center of the stars. To allow the series to reach the surface of the polytropic sphere and consequently to the surface of the star, we used the acceleration technique proposed by [7].

The first step is to compare and check the accuracy of the zeroth calculated (this value is equivalent to the radius of the star) from Equation (72) with that of the exact solution presented by Equations (29-31). We used the code developed by [16] to calculate the zeroth of the Emden function ($x_1$) at different fractional parameters. Tables (1-2) illustrate the results for the polytopes with n=0 and n=1 respectively. The third column is the zeroth computed by Equation (26) for the series expansion, with the aid of the recurrence relation, Equations (27) and (28) and the second column represents the zeroth computed from the exact solution of Equations (29-31). As is shown in the table, the maximum relative error is about 1.6 %.

The stellar mass-radius relation could be computed using Equations (32-33) and the series expansion for the first fractional derivative of Equation (72) is listed in Table (3) for mass-radius relation of the fractional polytrope with n=3 [16]. In this table, the ratio $R_*/R_0$ is the ratio of the radius of the star to the radius of the sun and $M_*/M_0$ is the ratio of the star mass to the solar mass.

Table 1: Radius of convergence for the n = 0 fractional polytropes.

| $\alpha$ | $x_1$ (exact) | $x_1$ (series) | Absolute relative error |
|---|---|---|---|
| 1 | 2.44 | 2.44 | 0 % |
| 0.99 | 2.424 | 2.435 | 0.45 % |
| 0.98 | 2.400 | 2.415 | 0.62 % |
| 0.97 | 2.376 | 2.405 | 1.2 % |
| 0.96 | 2.351 | 2.385 | 1.27 % |
| 0.95 | 2.327 | 2.365 | 1.63 % |

Table 2: Radius of convergence for the n = 1 fractional polytrope.

| $\alpha$ | $x_1$ (exact) | $x_1$ (series) | Absolute relative error |
|---|---|---|---|
| 1 | 3.14 | 3.14 | 0 % |
| 0.99 | 3.110 | 3.114 | 0 % |
| 0.98 | 3.078 | 3.085 | 0.23 % |
| 0.97 | 3.047 | 3.054 | 0.23 % |
| 0.96 | 3.015 | 3.035 | 0.7 % |
| 0.95 | 2.984 | 3.0 | 0.53 |

Table 3: Mass-radius relation for fractional polytrope with n=3 [16].

| $\alpha$ | $R_* / R_0$ | $M_* / M_0$ |
|---|---|---|
| 1 | 1 | 1 |
| 0.99 | 0.969 | 0.950 |
| 0.98 | 0.951 | 0.909 |
| 0.97 | 0.933 | 0.874 |
| 0.96 | 0.915 | 0.840 |
| 0.95 | 0.897 | 0.809 |

### 5.2 Network training

The training phase of the proposed neural network is implemented by computing the distributions of the Emden functions and mass-radius relation for the values listed in the second column of Tables (4-5).

Table 4: Training and testing data for the polytrope.

|   | Training phase | Testing phase |
|---|---|---|
| n | α | α |
| 0 | 0.96, 0.97, 0.98, 0.99, 1 | 0.95 |
| 1 | 0.95, 0.97, 0.98, 0.99, 1 | 0.96 |
| 3 | 0.95, 0.96, 0.98, 0.99, 1 | 0.97 |
| 5 | 0.95, 0.96, 0.98, 0.99, 1 | 0.97 |

Table 5: Training and testing data for mass-radius and density-radius relations.

|   | Training phase | Testing phase |
|---|---|---|
| n | α | α |
| 3 | 0.95, 0.98, 0.99, 1 | 0.96 |

The neural network (NN) used in this article uses two different configurations. For the polytropic case, the input layer of the NN has three individual inputs which are the polytropic index $n$, the fractional parameter $\alpha$ and the dimensionless parameter $x$ ($x$ takes values from 0 to $x_1$, where $x_1$ is the first zero of the Emden function as listed in Tables 1-2), while the output layer has 1 node for the Emden function $u$ calculated for the same values of the input fractional parameter and dimensionless parameter $x$. For the mass-radius relation with a polytropic index n=3 and various fractional parameters $\alpha$, the input layer of the NN has two individual inputs which are the radius and mass of the star, whereas the output layer has 2 nodes which are the radius and mass at the same values of the input fractional parameters.

After testing different configurations of hidden neurons of 80,120 and 200 neurons in the NN (shown in Figure 1), it was found that one hidden layer containing 120 neurons gives the best network model to compute accurately the exact fractional values of the Emden function. This is shown in Figure (3) and Figure (4) below for both polytropic and mass-radius relation cases respectively. In Figure (3), it is clear that the 120 neurons in the hidden layer case are giving the least RMS error compared to the other two configurations along the whole cycle of NN training cycles. The same remark is applied for Figure (4) for mass-radius relation case in which we can see oscillations for the RMS errors during NN training cycles after which the error for the 120 neurons case decreases to its final value better than the other 2 configurations.

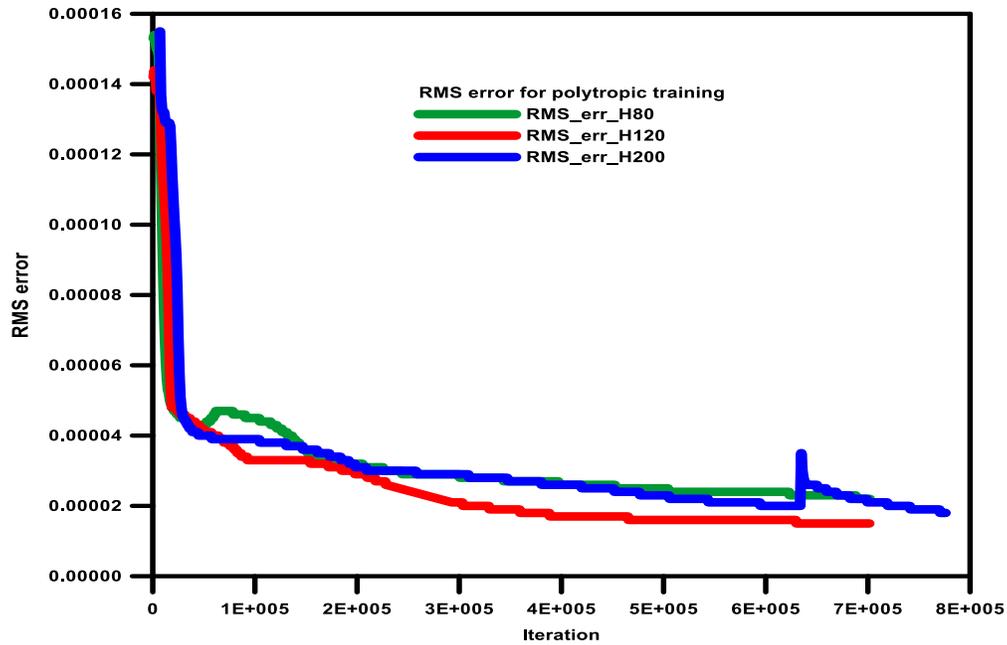

**Figure 3. RMS errors for a different number of hidden layer neurons for the polytropic case.**

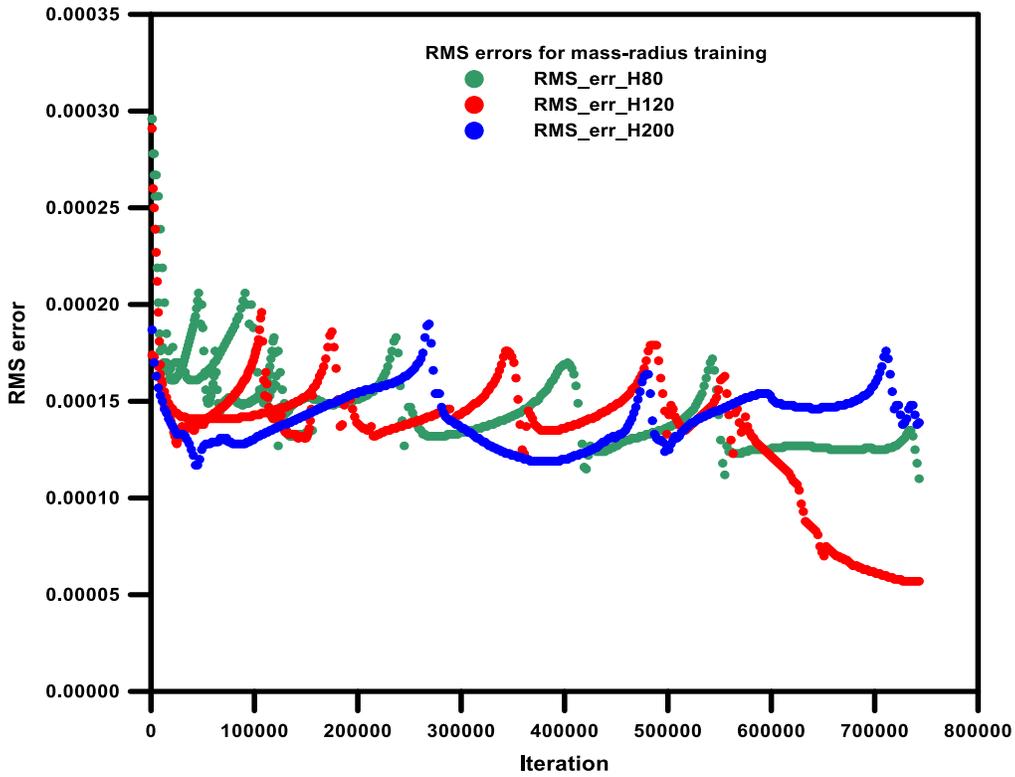

**Figure 4. RMS errors for a different number of hidden layer neurons for the case of mass-radius relation.**

After various adjustments and modifications to the network parameters, the network converged to a threshold RMS error of 0.000015 for the polytrope training case and of 0.000057 for the mass-radius relation training case. During those training, we used values of $\eta = 0.03$ for the learning rate and $\gamma = 0.5$ for the momentum. Those values for $\eta$ and $\gamma$ were found to speed up the convergence of the back-propagation learning algorithm of our ANN without over-shooting the solution. In order to demonstrate the stability and convergence of the computed values of weight parameters of the layers of the network, the convergence behaviors of the weights of the input layer, bias and the weights of output layer ( $w_i$, $\beta_i$ and $v_i$ ) for the polytropic case are displayed in Figure (5). Similarly, the stability and convergence behaviors of the computed values of weight parameters of the layers of the network for the input layer weights, bias, and output layer weights for the mass-radius relation case are displayed in Figure (6). As is shown in these figures, the weight values are initialized to some random values where they converge to stable values after somewhat large iteration values.

### 5.3 Comparison with exact and approximate solutions

By the end of the training phase of the ANN, its algorithm is ready to compute the Emden functions for the polytropic indices and the fractional parameters (third column) listed in Table (4). The results for the fractional polytropes are illustrated in Figures (7-10) for the following pairs of the polytropic indices and fractional parameters, (n=0, α=0.95), (n=1, α=0.96), (n=3, α=0.97) and (n=5, α=0.97). For n=0, 1, the Emden functions are computed using the exact solutions (Equations (29-31)), where for the polytropic index n=3 and due to the lack of the exact solution, the series solution is considered. To achieve the accuracy of the calculations, the ANN and the series solutions are plotted in the figures with different colors. As it is clear, they overlap and cannot be distinguished.

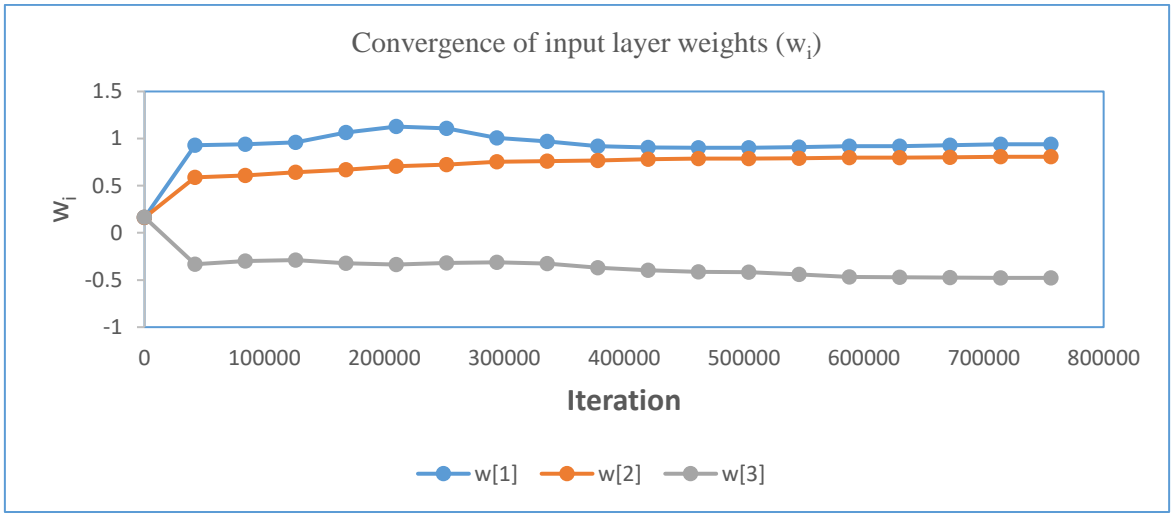

(a) The convergence of weights of the input layer ($w_i$)

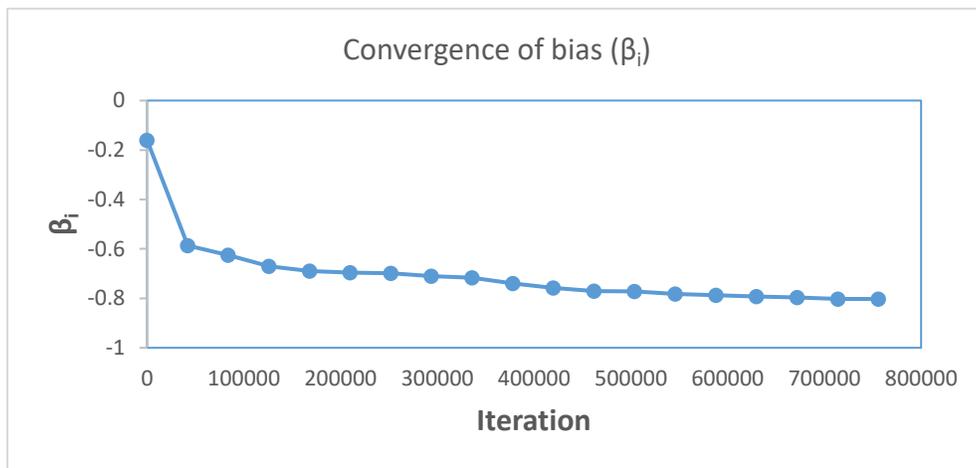

(b) The convergence of bias ($\beta_i$)

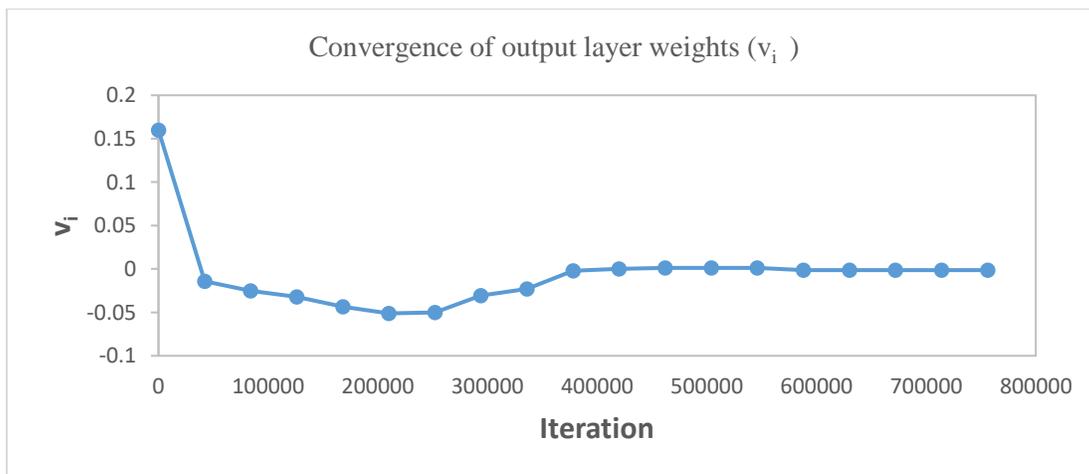

(c) The convergence of weights of output layer ($v_i$)

**Figure (5) Convergence of input, bias and output weights for the polytropic case**

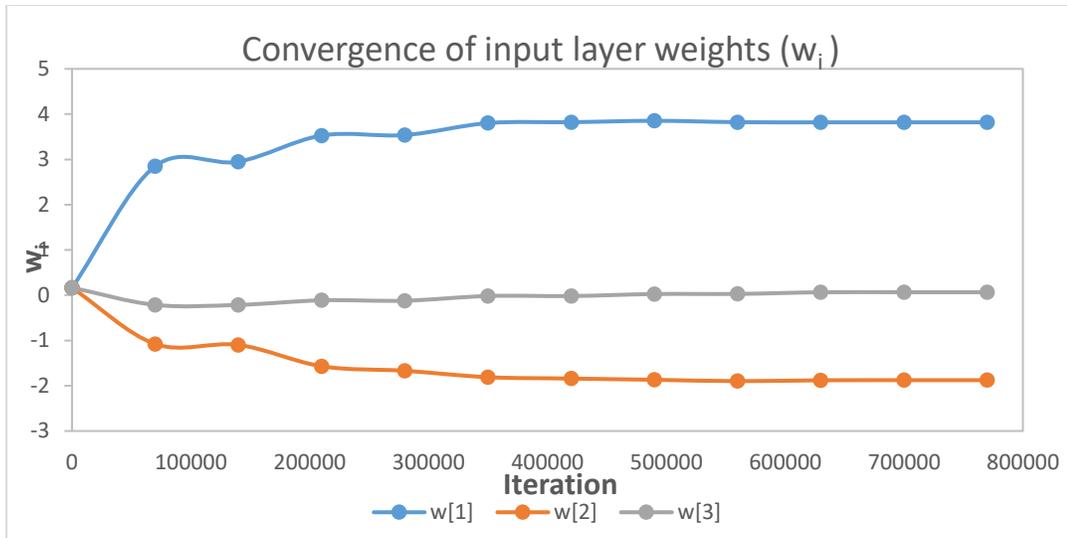

(a) The convergence of weights of the input layer ($w_i$)

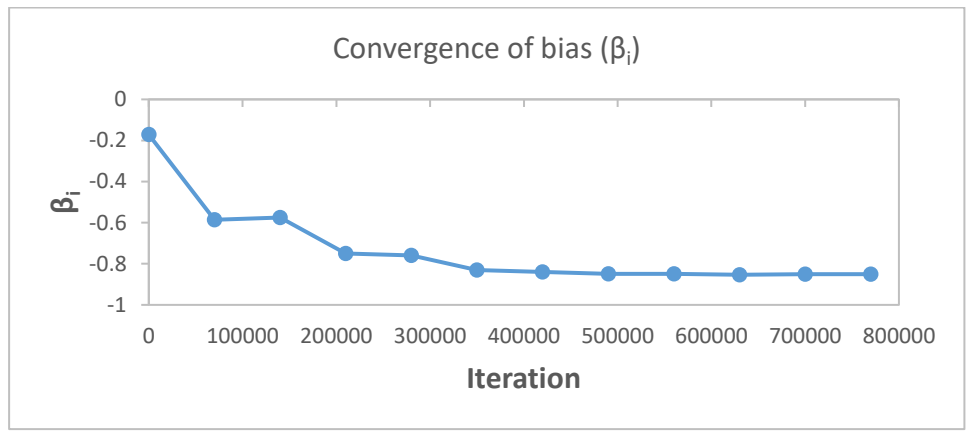

(b) The convergence of bias ($\beta_i$)

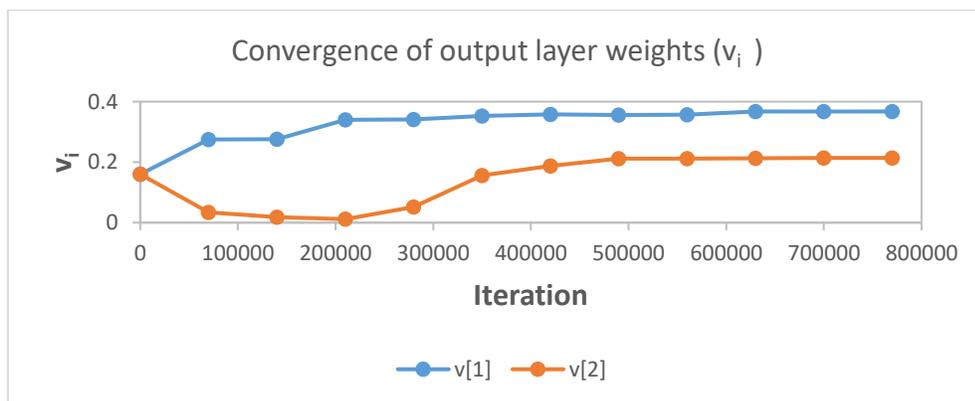

(c) The convergence of weights of output layer ($v_i$)

**Figure (6) Convergence of input, bias and output weights for the mass-radius relation case**

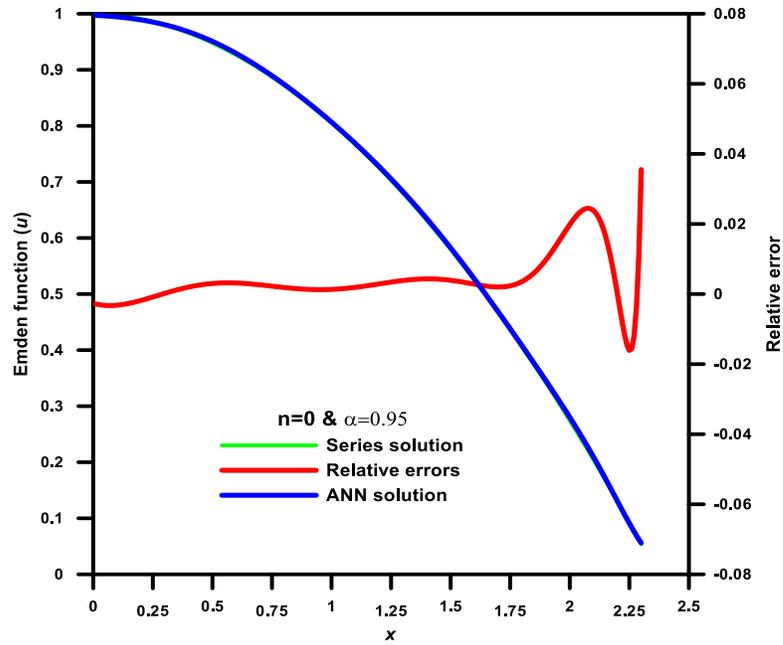

**Figure 7: The fractional Emden function of the polytrope with n=0 and $\alpha$=0.95. The maximum relative error is 3.5%.**

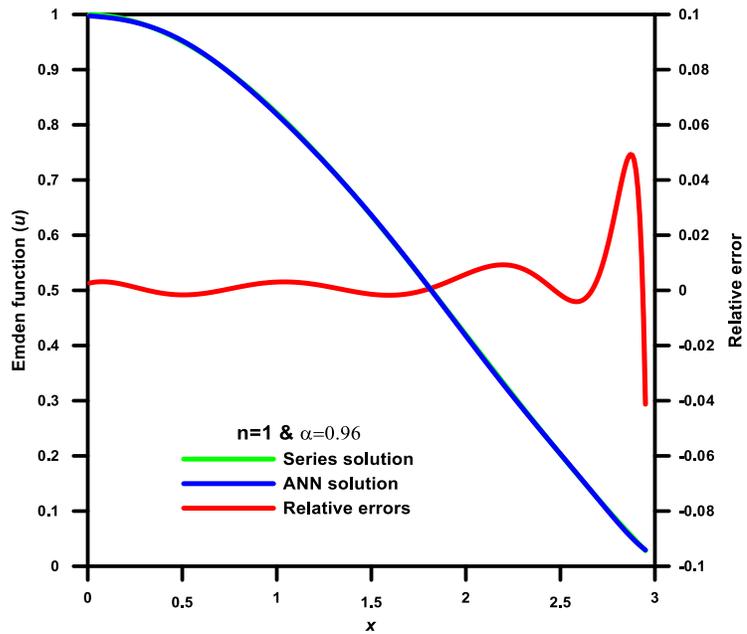

**Figure 8: The fractional Emden function of the polytrope with n=1 and $\alpha$=0.96. The maximum relative error is 5%.**

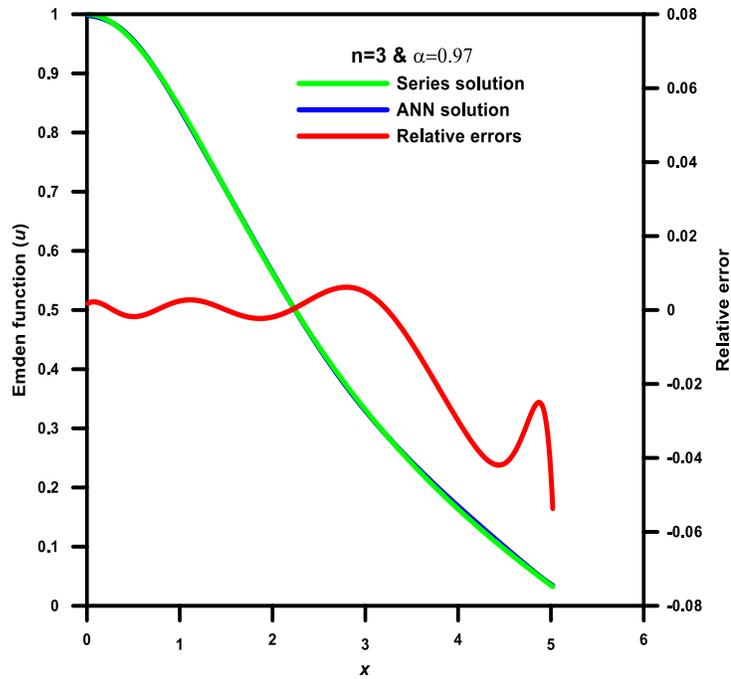

**Figure 9: The fractional Emden function of the polytrope with n=3 and α=0.97. The maximum relative error is 5.4%.**

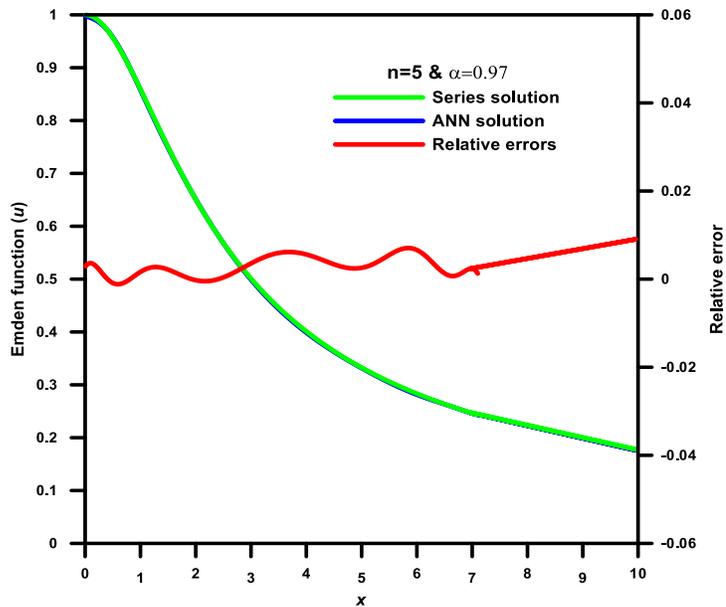

**Figure 10: The fractional Emden function of the polytrope with n=5 and α=0.97. The maximum relative error is 1%.**

The maximum relative errors are, 3.5%, 5%, 5.4%, 1% respectively. The large errors that are appeared in some regions in the curves may be attributed to computer accuracy.

In astrophysics, observational verification of the theoretical mass-radius relationship was a prime objective of numerous studies that considered individual stars and stellar associations with strong determinations of mass and radius individual stars and radius [51]. Following this motivation, we tried to model the mass-radius relation of the fractional polytrope with n=3. Figure (11) displays the series and the ANN models of the mass-radius relations. As we see, the two models are generally in good agreement except for the middle part of the distribution. One of the computational difficulties that may cause this numerical instability when computing the mass of the polytrope, is the fractional derivative appeared in Equation (32).

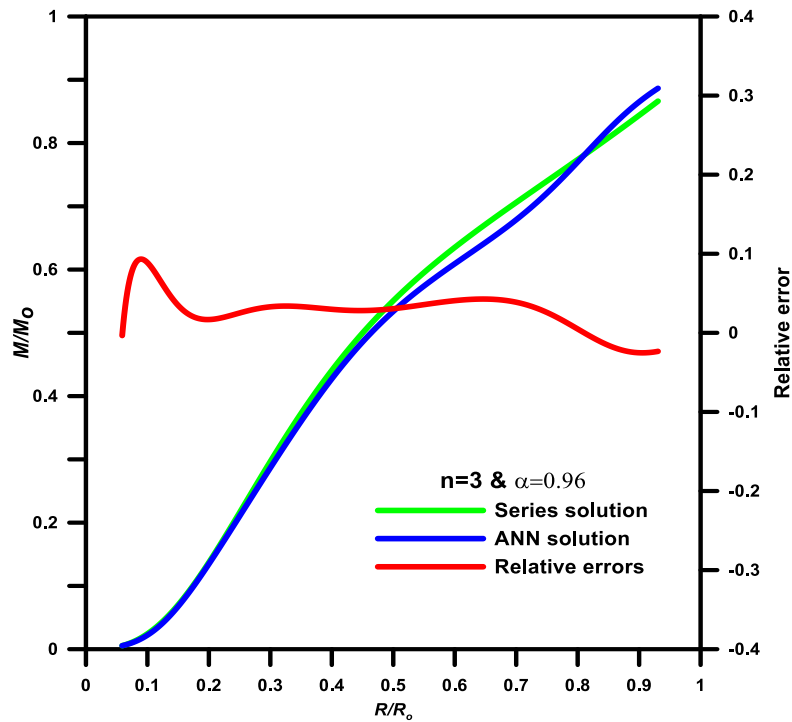

**Figure 11: The fractional mass-radius relation for the polytropic star with n=3 and α=0.96. The maximum relative error is 9%.**

## 6. Conclusions

In this paper, we introduced an artificial neural network approach for solving the fractional Lane-Emden equation of the polytropic gas spheres. We used the ANN in its feedforward back propagation learning scheme. The input data for the training phase and that for testing were created using the code developed by [16], where the analytical solution is performed using the series expansion method. We have predicted the distribution of the Emden functions for the polytropic indices having exact solutions, n=0, 1, 5. Also, the approach was successfully applied to model the mass-radius relation of the polytropic star with n=3. The results reached, had been compared with the exact solution as well as the series expansion solution. These results show that the ANN and the series solutions can barely be distinguished; which consequently means that the neural network works very well in predicting values of fractional Lane-Emden functions and small errors were found in the outputs. A possible weakness of the obtained results may arise from the nature of the ANN method; it somewhat more rigid than the numerical and the analytical methods used to solve the fractional Lane-Emden equation. When a neural network is trained in looking for all the parameters, it requires to learn in a version of the question that can use additional knowledge.

**Conflict of Interest:** The authors declare that they have no conflict of interest.